\documentclass[journal]{IEEEtran}

\usepackage{graphicx}
\usepackage{amsmath}
\usepackage[version=4]{mhchem}
\usepackage{siunitx}
\usepackage{longtable,tabularx}
\usepackage{color}

\usepackage{amssymb}

\usepackage{algorithm} 
\usepackage{algpseudocode} 

\newtheorem{assumption}{Assumption}
\newtheorem{theorem}{Theorem}
\newtheorem{definition}{Definition}
\newtheorem{remark}{Remark}
\newtheorem{example}{Example}

\begin{document}
    \title{Dynamic System Stability Verification Using  Numerical Simulator}
    \author{Jongrae~Kim
    \\
    \thanks{ 
    This work was funded by Unmanned Vehicles Core Technology Research and  Development Program (No.\,2020M3C1C1A0108316111) through the National Research Foundation of Korea (NRF) and Unmanned Vehicle Advanced Research Center(UVARC) funded by the Ministry of Science and ICT, Republic of Korea.}
    \thanks{J.~Kim is with the School of Mechanical Engineering, University of Leeds, Leeds LS2 9JT, UK (e-mail: menjkim@leeds.ac.uk).}
}  

\maketitle

\begin{abstract}
There are recent shifts in demand for design controllers from simplified to complex model-based. Although simplification approaches are successful in many areas of engineering control systems, high-fidelity simulation-based control design, for example, reinforcement learning, has been rising in robotics areas. On the other hand, the lack of assurances about the stability and robustness of simulation-based control design restricts its applications to safety-critical systems.
We develop computational methods to verify the stability
and robustness of safety-critical systems. By extending the inverse
Lyapunov theorem, we present a practical method to compute the
constants required to check the exponential stability conditions of dynamic systems implemented in a numerical simulator. It is shown that the norm-bound of the propagated states is a function of the numerical integration steps, where
the numerical simulator may include discontinuous
jumps of states. The energy bounds for the transition states
are obtained based on the exponential stability assumption of the
inverse Lyapunov theorem. Finally, a finite sampling algorithm provides the deterministic stability guarantee for the continuous state space.   
\end{abstract}

\section{Introduction}
Most control design approaches rely on simplified dynamic model descriptions for complex real-world systems. These simplification approaches have been very successful in many engineering systems over the past few decades \cite{1102565,29425,1101446,morari1999model}. However, all control system designs must undergo costly, laborious, and tedious system verification procedures through computational \cite{7741019} or experimental methods \cite{5340639,8723561}.
While there are several immediate challenges in implementing accurate and computationally efficient numerical simulators, it is a common practice to use high-fidelity simulators to verify the performance and robustness of controllers for many engineering systems.
In control system design projects, frequently high-fidelity dynamic simulators are available. For example, combining the rotor and wing models for a compound aircraft simulator is presented in \cite{lee2021development}. A detailed quad-copter vehicle simulator for abnormal simulations including the full rigid-body dynamics, the propulsion model, the aerodynamics model and the low-level controller has been shown in \cite{foster2017high}. Design procedures for implementing a simulator model of an electric motor 
producing the same responses as the real motor is demonstrated in \cite{hieb2013creating}.

Meanwhile, several simulation-based control design approaches have been on the rise recently \cite{sutton1999reinforcement, lillicrap2015continuous}. The rise of reinforcement learning to solve challenging control problems is particularly noticeable in robotics. The contact dynamics of robot manipulators are difficult to take into account explicitly in the control design steps \cite{vukobratovic2003dynamics}. The nature of simulation-based control design approaches of reinforcement learning makes it the ideal tool for the control design. On the other hand, the lack of assurances about the stability and robustness of simulation-based control systems makes it challenging to deploy the designs in safety-critical systems such as aircraft, spacecraft and rockets.

In the following sections, we derive the norm bound of states propagated
by numerical simulators. The exact calculation of the norm bounds is 
critical for applying the converse Lyapunov theorem in the stability
verification. Based on the norm condition with the exponential stability
assumption, a stability assurance algorithm using a finite
number of simulations providing the deterministic stability assurance
is presented. Finally, the conclusions and future works are discussed.

\section{Norm-bound of simulator propagated states}
Consider the nonlinear system given by \cite{khalil2015nonlinear}
\begin{align} \label{eq:dxdt_f_x}
	\dot{\bf x} &= {\bf f}(t, {\bf x}) 
\end{align}
where
${\bf x}$ is a real-valued $n$-dimensional vector belong to ${\mathbb R}^n$, 
which is the $n$-dimensional  real space, 
$\dot{\bf x}=d(\cdot)/dt$ is the derivative of ${\bf x}$ with respect to the time, $t$,
and ${\bf f}(t,{\bf x})$ is a continuously differentiable nonlinear function with respect to ${\bf x}$ in 
${\mathbb D} = \{  {\bf x} | \|{\bf x} \| < r \}$ for $r > 0$.  ${\bf f}(t, {\bf x})$  would include a feedback control system
and $\partial{\bf f}/\partial {\bf x}$ is bounded on ${\mathbb D}$ implying that $\|{\bf f}(t,{\bf x})\|$ is bounded. 
In addition, for the uniqueness of the solution, the Lipschitz condition must be satisfied:
$\| {\bf f}(t,{\bf x}) - {\bf f}(t,{\bf y})\| \le L \| {\bf x} - {\bf y} \|$ for any ${\bf x}$ and ${\bf y}$ in ${\mathbb D}$
and $L > 0$.
The continuous differentiability and Lipschitz continuity requirements are restrictive 
for engineering systems to satisfy without introducing tight bounds on
the operational conditions. 
\bigskip

Instead of the differential equation form of a nonlinear system, \eqref{eq:dxdt_f_x}, 
consider the following integral-type nonlinear systems
\begin{align} \label{eq:x_t_x0_int_f}
	{\bf x}(t) = {\bf x}(t_0) + \int_{\tau = t_0}^{\tau = t} {\bf f}[t, {\bf x}(\tau)] d\tau
\end{align}
where ${\bf f}(t, {\bf x})$ is Lebesgue integrable, which allows discontinuous jumps in 
the finite number of isolated instances, $t$. 
This is called the Carath\'{e}odory solution \cite{trumpf2010converse}.
By the definition of \eqref{eq:x_t_x0_int_f}, the states are automatically continuous and
continuously differentiable for almost all $t$.
The Lipschitz condition is again required to be satisfied. 
This is a less restrictive description of nonlinear systems
than \eqref{eq:dxdt_f_x}, but it still does not allow discontinuous jumps of ${\bf f}(t,{\bf x})$ 
in ${\bf x}$.

\bigskip

From now on, we consider the systems without the explicit dependence on time such that 
\eqref{eq:x_t_x0_int_f} becomes
\begin{align} \label{eq:x_t_x0_int_f_no_time_dep}
{\bf x}(t) = {\bf x}(t_0) + \int_{\tau = t_0}^{\tau = t} {\bf f}[{\bf x}(\tau)] d\tau
\end{align}
where the continuity of ${\bf f}[{\bf x}(t)]$ in ${\bf x}(t)$ is to be relaxed later.
The numerical implementation of \eqref{eq:x_t_x0_int_f_no_time_dep} is
\begin{align} \label{eq:x_t_x0_sum_int_f}
	{\bf x}(t+\Delta t) 
    &= \Phi[t+\Delta t, t, {\bf x}(t)] \notag\\
    &= {\bf x}(t) 
	+ \int^{\tau = t+\Delta t}_{\tau = t} {\bf f}[{\bf x}(\tau)] d\tau
\end{align}
where $\Phi(t_1, t_0, {\bf x}_0)$ is the state transition function from $t_0$ to $t_1$ starting
at the initial state, ${\bf x}_0$ at $t_0$ and the integral in the right-hand side is implemented by
a numerical method such as the Euler method, i.e.,
\begin{align} \label{eq:euler_integral}
	{\boldsymbol\Phi}_{\text{Euler}}[t+\Delta t, t, {\bf x}(t)] = {\bf x}(t) + {\bf f}[{\bf x}(t)] \Delta t
\end{align}
or the Runge-Kutta method, i.e.,
\begin{align}\label{eq:rk_integral}
	{\boldsymbol\Phi}_{\text{RK}}[t+\Delta t, t, {\bf x}(t)] &= {\bf x}(t) \notag\\
      &+ \dfrac{\Delta t}{6}
		\left( {\bf k}_1 + 2 {\bf k}_2 + 2 {\bf k}_3 + {\bf k}_4 \right)
\end{align}
where
\begin{align*}
	{\bf k}_1[\Delta t, {\bf x}(t)] &= {\bf f}[{\bf x}(t)]\\
    {\bf k}_2[\Delta t, {\bf x}(t)] &= {\bf f}[{\bf x}(t) + {\bf k}_1 \Delta t/2]\\
    {\bf k}_3[\Delta t, {\bf x}(t)] &= {\bf f}[{\bf x}(t) + {\bf k}_2 \Delta t/2]\\
    {\bf k}_4[\Delta t, {\bf x}(t)] &= {\bf f}[{\bf x}(t) + {\bf k}_3 \Delta t]
\end{align*}
As the numerical solution of \eqref{eq:x_t_x0_sum_int_f}, i.e., \eqref{eq:euler_integral}
or \eqref{eq:rk_integral}, given by the numerical simulators is frequently the only available result in practice.

\bigskip

\begin{example}[Numerical Solution for Non-Lipschitz Systems]
    The nonlinear system, $\dot{x} = f(x)$, where $f(x)= x^{1/3}$ and $x(0) = 0$,
    has two solutions, i.e., $x(t)=0$ and $x(t) = (2t/3)^{3/2}$ \cite{khalil2015nonlinear}.
    The cause of the multiple solutions is that the slope of $x^{1/3}$ at $x=0$ is infinity.
    Hence, the Lipschitz condition restricts the slope of $f(x)$ so that
    the uniqueness of the solution is guaranteed. On the other hand, applying the Euler
    or the Runge-Kutta integration to the nonlinear system returns $x(t)=0$ as the solution.
\end{example}

\bigskip

\begin{theorem}[Lipschitz Condition for ${\boldsymbol\Phi}_\text{\rm Euler}$ and ${\boldsymbol\Phi}_\text{\rm RK}$]
\label{theorem:lipschitz_euler_rk}
	For any ${\bf x}$ and ${\bf y}$ in ${\mathbb D}$, if ${\bf f}({\bf x})$ satisfies
	\begin{align}
		\| {\bf f}({\bf x}) - {\bf f}({\bf y})\| \le L \| {\bf x} - {\bf y} \|
	\end{align}
	then
	\begin{align}
		&\| {\boldsymbol\Phi}_\text{\rm Euler}[t+\Delta t, t, {\bf x}] 
		- {\boldsymbol\Phi}_\text{\rm Euler}[t+\Delta t, t, {\bf y}]\|
		\notag\\
        &\le
		\left( 1 + L \Delta t \right)
		\| {\bf x} - {\bf y} \|
	\end{align}
	and
	\begin{align}
		&\| {\boldsymbol\Phi}_\text{\rm RK}[t+\Delta t, t, {\bf x}] 
		- {\boldsymbol\Phi}_\text{\rm RK}[t+\Delta t, t, {\bf y}]\|
		\\
        &\le
		\left[
		1 + L \Delta t + \dfrac{(L \Delta t)^2}{2} 
		+ \dfrac{(L \Delta t)^3}{6} + \dfrac{(L \Delta t)^4}{24}
		\right] 
		\| {\bf x} - {\bf y} \| \notag
	\end{align}
\end{theorem}
\emph{Proof:} The proof is straightforward and omitted or 
see the proof of Theorem \ref{theorem:bound_euler_rk}. 
\begin{flushright}
$\blacksquare$    
\end{flushright}

\bigskip

\begin{remark} (Conservatism of the bounds for ${\boldsymbol\Phi}_\text{\rm Euler}$ 
	and ${\boldsymbol\Phi}_\text{\rm RK}$)
Given the Lipshitz condition of ${\bf f}({\bf x})$ in ${\mathbb D}$, the transition
functions also satisfy the Lipschitz condition with a Lipschitz constant being a function of $\Delta t$.
The bounds in Theorem \ref{theorem:lipschitz_euler_rk}
cannot be arbitrarily small by decreasing $\Delta t$ as it is the simulation marching step in time. 
$\Delta t$ equal to zero
provides the trivial transition function, i.e., the identity.
\end{remark}
\bigskip

The function implemented in a high-fidelity computer simulator, 
${\bf f}({\bf x})$ in \eqref{eq:x_t_x0_sum_int_f},
typically includes nonlinear and discontinuous components such as saturation, 
friction, backlash, hysteresis, deadband and so forth. 
Given that numerically solving the dynamic simulator expressed in mathematical form as 
\eqref{eq:dxdt_f_x} or equivalently
implementing a numerical integration for \eqref{eq:x_t_x0_int_f} imposes a different condition on ${\bf f}({\bf x})$.
\bigskip

\begin{theorem}[Bound for ${\boldsymbol\Phi}_\text{\rm Euler}$ and ${\boldsymbol\Phi}_\text{\rm RK}$]
\label{theorem:bound_euler_rk}
	For any ${\bf x}$ and ${\bf y}$ in ${\mathbb D}$, if ${\bf f}({\bf x})$ satisfies
	\begin{align} \label{eq:f_bound_L_and_M}
        \| {\bf f}({\bf x}) - {\bf f}({\bf y})\| \le L \| {\bf x} - {\bf y} \| + M 
	\end{align}
	where $M$ greater than or equal to zero is the maximum possible discontinuity of 
	${\bf f}({\bf x})$, 
	then
    \begin{align} \label{eq:phi_el_bound}
		&\| {\boldsymbol\Phi}_\text{\rm Euler}[t+\Delta t, t, {\bf x}] 
		- {\boldsymbol\Phi}_\text{\rm Euler}[t+\Delta t, t, {\bf y}]\|
		\notag\\
        &\le
        \left( 1 + L \Delta t \right) \| {\bf x} - {\bf y} \| +  \Delta t M
	\end{align}
	and
	\begin{align} \label{eq:phi_rk_bound}
		&\| {\boldsymbol\Phi}_\text{\rm RK}[t+\Delta t, t, {\bf x}] 
		- {\boldsymbol\Phi}_\text{\rm RK}[t+\Delta t, t, {\bf y}]\|
		\notag\\
        &\le
		( 1 + L \alpha) \| {\bf x} - {\bf y} \| + \alpha M
	\end{align}
	where
	\begin{align}
        \alpha = \Delta t \left[ 1 + \dfrac{L \Delta t}{2} 
        + \dfrac{(L \Delta t)^2}{6} + + \dfrac{(L \Delta t)^3}{24} \right]
	\end{align}
\end{theorem}
\emph{Proof:} For the Euler integral, the proof is straightforward. In the following, we show
the proof for the Runke-Kutta integral.
\begin{align*}
    \Delta {\bf k}_1 
        &= \| {\bf k}_1(\Delta t, {\bf x}) - {\bf k}_1(\Delta t, {\bf y}) \|
        \notag\\
        &= \| {\bf f}({\bf x}) - {\bf f}({\bf y}) \| \le L \| {\bf x} - {\bf y} \| + M
\end{align*}
\begin{align*}
    \Delta {\bf k}_2 
        &= \| {\bf k}_2(\Delta t, {\bf x}) - {\bf k}_2(\Delta t, {\bf y}) \|
        \notag\\
        &= \| {\bf f}[{\bf x} + {\bf k}_1(\Delta t, {\bf x}) \Delta t/2] 
           - {\bf f}[{\bf y} + {\bf k}_1(\Delta t, {\bf y}) \Delta t/2] \|\\
        &\le L \| {\bf x} + {\bf k}_1(\Delta t, {\bf x}) \Delta t/2 
                - {\bf y} - {\bf k}_1(\Delta t, {\bf y}) \Delta t/2\| + M\\
        &\le L \left( 1 + \dfrac{L \Delta t}{2} \right) \| {\bf x} - {\bf y} \|
        + \left( 1 + \dfrac{L\Delta t}{2} \right) M
\end{align*}
\begin{align*}
    \Delta {\bf k}_3 
        &= \| {\bf k}_3(\Delta t, {\bf x}) - {\bf k}_3(\Delta t, {\bf y}) \|
        \notag\\
        &= \| {\bf f}[{\bf x} + {\bf k}_2(\Delta t, {\bf x}) \Delta t/2] 
           - {\bf f}[{\bf y} + {\bf k}_2(\Delta t, {\bf y}) \Delta t/2] \|\\
        &\le L \| {\bf x} + {\bf k}_2(\Delta t, {\bf x}) \Delta t/2 
                - {\bf y} - {\bf k}_2(\Delta t, {\bf y}) \Delta t/2\| + M\\
        &\le L \left( 1 + \dfrac{L \Delta t}{2} + \dfrac{L^2 \Delta t^2}{4}\right) \| {\bf x} - {\bf y} \|\\ 
        &~~~+ \left( 1 + \dfrac{L\Delta t}{2} + \dfrac{L^2 \Delta t^2}{4} \right) M
\end{align*}
and
\begin{align*}
    \Delta {\bf k}_4 
        &= \| {\bf k}_4(\Delta t, {\bf x}) - {\bf k}_4(\Delta t, {\bf y}) \|\\
        &= \| {\bf f}[{\bf x} + {\bf k}_3(\Delta t, {\bf x}) \Delta t] 
           - {\bf f}[{\bf y} + {\bf k}_3(\Delta t, {\bf y}) \Delta t] \|\\
        &\le L \| {\bf x} + {\bf k}_3(\Delta t, {\bf x}) \Delta t 
                - {\bf y} - {\bf k}_3(\Delta t, {\bf y}) \Delta t\| + M\\
        &\le L \left( 1 + L \Delta t + 
        \dfrac{L^2 \Delta t^2}{2} + \dfrac{L^3 \Delta t^3}{4}\right) \| {\bf x} - {\bf y} \|\\ 
        &~~~+ \left( 1 + L \Delta t + \dfrac{L^2 \Delta t^2}{2} + \dfrac{L^3 \Delta t^3}{4} \right) M
\end{align*}
Apply the above inequalities to the following inequality:
	\begin{align}
        &\| {\boldsymbol\Phi}_\text{\rm RK}[t+\Delta t, t, {\bf x}(t)] 
		- {\boldsymbol\Phi}_\text{\rm RK}[t+\Delta t, t, {\bf y}(t)]\|
		\notag\\
        &\le
        \| {\bf x} - {\bf y} \| + \dfrac{\Delta t}{6} \left(
                \Delta {\bf k}_1 + 2 \Delta {\bf k}_2 + 2 \Delta {\bf k}_3 + \Delta {\bf k}_4
            \right)\notag\\
        &=
        \left[
		1 + L \Delta t \left( 1 + \dfrac{L \Delta t}{2} 
		+ \dfrac{(L \Delta t)^2}{6} + + \dfrac{(L \Delta t)^3}{24}
            \right)
		\right] 
		\| {\bf x} - {\bf y} \|
        \notag\\
        &~~~+
        \Delta t \left[
		1 + \dfrac{L \Delta t}{2} 
		+ \dfrac{(L \Delta t)^2}{6} + + \dfrac{(L \Delta t)^3}{24}
		\right] M
	\end{align}
\begin{flushright}
$\blacksquare$
\end{flushright}
\bigskip

\begin{example} \label{example:sgn_cubic}
	A nonlinear system is given by $\dot{x} = - 2 \text{\rm sgn}(x) + x^3/3$.
	The bound for ${\mathbb D}  = \left\{ x |~ |x| < 3/2 \right\}$ and
    $\Delta t = 0.01$ is obtained as follows:
	\begin{align*}
	| f(x) - f(y) | 
        &= \left| \left(- 2\text{\rm sgn}(x) + x^3/3 \right) 
	  - \left(- 2\text{\rm sgn}(y) + y^3/3 \right)
      \right|\\
        &\le 2 | \text{\rm sgn}(y) - \text{\rm sgn}(x) | 
        + \dfrac{1}{3}|x^3 - y^3|\\
        &\le 2 \times 2
    + \dfrac{1}{3} \times \left[~\max_{x \in {\mathbb D}} 
        \left|\dfrac{d(x^3/3)}{dx}\right| ~\right] |x-y|\\
        &= 2 \times 2
        + \dfrac{1}{3} \times \left(\dfrac{3}{2}\right)^2|x - y| = \dfrac{3}{4} |x-y| + 4
	\end{align*}
	i.e., 
    $L=3/4$ and $M = 4$. Therefore,
    \begin{align*}
    &\| {\boldsymbol\Phi}_\text{\rm RK}[t+\Delta t, t, x] 
		- {\boldsymbol\Phi}_\text{\rm RK}[t+\Delta t, t, y]\|
        \\
        &= 1.0075 \| x - y \| + 0.040
    \end{align*}
    For $\Delta t = 0.001$ or $0.5$, 
    $\alpha$  is 0.0010 or 0.606, 
    respectively.
\end{example}

\bigskip

\begin{definition}[State-Transition by Numerical Simulator]
The states propagated by the numerical simulator is given by
\begin{align} \label{eq:main_dsc_sys}
    {\bf x}(t+\Delta t) = {\boldsymbol\Phi}[t+\Delta t, t, {\bf x}(t)]
\end{align}
where ${\boldsymbol\Phi}$ is the numerical integrator, 
e.g., ${\boldsymbol \Phi}_\text{Euler}$ or ${\boldsymbol \Phi}_\text{RK}$, 
and 
${\boldsymbol\Phi}$ is bounded by \eqref{eq:phi_el_bound} or \eqref{eq:phi_rk_bound}.
\end{definition}
\bigskip

\begin{assumption}[Existence \& Uniqueness of the Solution]
There is no unique way to define the solution of the nonlinear system given by 
\eqref{eq:x_t_x0_int_f_no_time_dep} with the discontinuous function bounded by \eqref{eq:f_bound_L_and_M}.
A good tutorial about various approaches to the solution is found in \cite{4518905}. 
We assume that the category of nonlinear systems considered here has a unique solution. 
\end{assumption}
\bigskip

\begin{assumption}[Nonlinear Simulator]
The trajectory obtained by recursive calculations of ${\bf x}(t+\Delta t)$ 
using \eqref{eq:main_dsc_sys} is given by
${\boldsymbol \phi}_N(t_0+k \Delta t, t_0, {\bf x}_{0})$ for a positive integer $k$, where $t_0$ is the initial time
and ${\bf x}_0 = {\bf x}(t_0)$ is the initial condition.
${\boldsymbol \phi}_N(t_0+ k \Delta  t, t_0, {\bf x}_{0})$ can be made
sufficiently close to the true solution,
${\boldsymbol \phi}(t_0 + k \Delta t, t_{0}, {\bf x}_{0})$, for
all positive integers, $k$.
\end{assumption}

\bigskip

\begin{theorem}[Bound for Longer Propagation]
The state transition from ${\bf x}(t)$ to ${\bf x}(t+T)$, where
$T$ is equal to $N \Delta t$ and $N$ is a positive integer, is given by
\begin{align} 
    &{\bf x}(t+T) = {\boldsymbol\phi}_N(N \Delta t, t, {\bf x})
    \\
    &= {\boldsymbol\Phi}[t+N\Delta t,t+(N-1)\Delta t, {\bf x}(t+(N-1)\Delta t)] \circ
    \ldots \notag\\
    &\ldots \circ {\boldsymbol\Phi}[t+2 \Delta t,t+\Delta t, {\bf x}(t+\Delta t)] \circ
    {\boldsymbol\Phi}[t+\Delta t,t, {\bf x}(t)] \notag
\end{align}
where $\circ$ is the composition operator and ${\boldsymbol\Phi}$ is assumed to be
the Runge-Kutta integral. The composition transfer function, 
${\boldsymbol\phi}_N(N \Delta t, 0, {\bf x})$, is bounded by
\begin{align} \label{eq:phi_bound_for_long_time_T}
 \| {\boldsymbol\phi}_N(N \Delta t, t, {\bf x}) - {\boldsymbol\phi}_N(N \Delta t, t, {\bf y}) \|
 \le a \| {\bf x} - {\bf y} \| + b
\end{align}
where
\begin{align}    
  a &= ( 1 + L \alpha)^N\\
  b &= \sum_{r=0}^{N-1} (1+L\alpha)^r \alpha M
\end{align}
\end{theorem}
\emph{Proof:} The bound for the time interval equal to $[0, 2 \Delta t]$ is given by
\begin{align}
	&\| {\boldsymbol\phi}_N(t+2\Delta t, t, {\bf x}) 
    -  {\boldsymbol\phi}_N(t+2\Delta t, t, {\bf y}) \|
    \notag\\ 
    &\le
    (1 + L \alpha) \| {\boldsymbol\phi}_N(t+\Delta t, t, {\bf x}) 
    -  {\boldsymbol\phi}_N(t+\Delta t, t, {\bf y}) \| + \alpha M  
    \notag \\
    &\le
    (1 + L \alpha) \left[ (1 + L \alpha) \|{\bf x} - {\bf y}\| + \alpha M  \right] + \alpha M
    \notag \\
    &= (1 + L \alpha)^2 \|{\bf x} - {\bf y}\| 
    + \left[ (1 + L \alpha) + 1\right] \alpha M
\end{align}
Similarly, the bound for the time interval equal to $[0, 3 \Delta t]$ is given by
\begin{align}
	&\| {\boldsymbol\phi}_N(t+3\Delta t, t, {\bf x}) 
    -  {\boldsymbol\phi}_N(t+3\Delta t, t, {\bf y}) \| 
    \notag\\
    &\le
    (1 + L \alpha) \| {\boldsymbol\phi}_N(t+2\Delta t, t, {\bf x}) 
    -  {\boldsymbol\phi}_N(t+2\Delta t, t, {\bf y}) \| + \alpha M 
    \notag\\
    &\le
    (1 + L \alpha)^3 \|{\bf x} - {\bf y}\| 
    + \sum_{\ell=0}^2 (1+L\alpha)^\ell \alpha M
\end{align}
By induction
\begin{align}
	&\| {\boldsymbol\phi}_N(t+N\Delta t, t, {\bf x}) 
    -  {\boldsymbol\phi}_N(t+N\Delta t, t, {\bf y}) \| 
    \notag\\
    &\le    (1 + L \alpha)^N \|{\bf x} - {\bf y}\| 
    + \sum_{r=0}^{N-1} (1+L\alpha)^r \alpha M
\end{align}
\begin{flushright}
$\blacksquare$    
\end{flushright}
\bigskip

\begin{figure}
    \centering
    \includegraphics[width=0.95\linewidth]{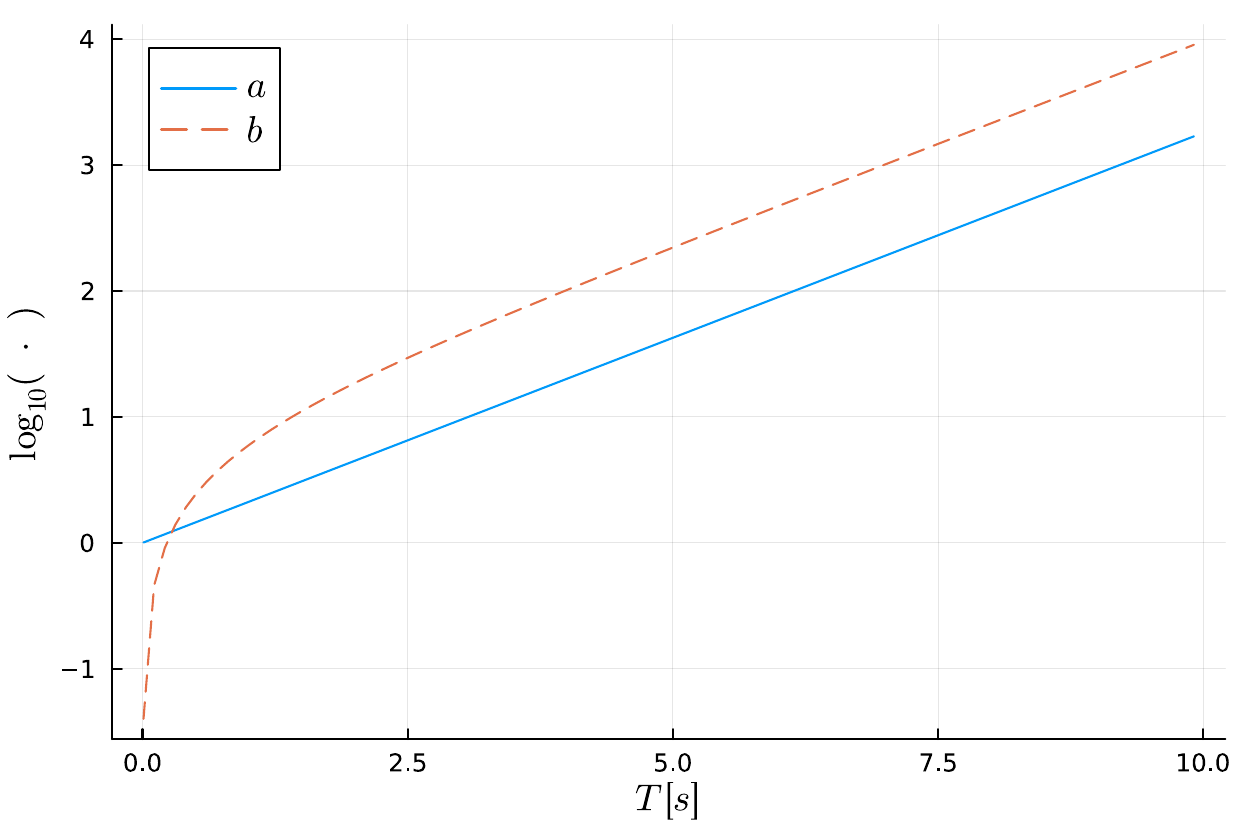}
    \caption{$\log(a)$ and $\log(b)$ with respect to $T$}
    \label{fig:example_a_and_b}
\end{figure}

\begin{example} \label{example:sgn_cubic_a_b_bounds}
For the system given in Example \ref{example:sgn_cubic}, 
where $(1+L\alpha) = 1.0075$, $\alpha M = 0.040$ and $\Delta t = 0.01$,
let the number of $\Delta t$, i.e., $N$, equal to 2,000 providing the simulation
time interval from 0 to $T$ equal to $20$s. Then,
\begin{align}
&\| {\boldsymbol\phi}_{2000}(20, 0, {\bf x}) 
- {\boldsymbol\phi}_{2000}(20, 0, {\bf y}) \|
\notag\\
 &\le
 1.0075^{2000} \| {\bf x} - {\bf y} \| + \sum_{r=0}^{1999} 1.0075^r \times 0.010
 \notag\\
 &\approx 3.27\times 10^6 \| {\bf x} - {\bf y} \| + 1.74\times 10^7 
\end{align}
The values of $a$ and $b$ in the bounds calculated are large. 
For $T$ from 1 to 10 seconds, Figure \ref{fig:example_a_and_b} shows their values. They become several hundred already around $T = 10$s, where $N=1000$.
\end{example}
\bigskip

In the following section, the bounds are improved by introducing the exponential stability assumption.

\section{Stability Verification}
To reduce the bounds obtained in the previous section, the exponential stability condition is introduced.
\begin{definition}[Exponential Stability]
    The equilibrium point, ${\bf x}_\text{eq}={\bf 0}$,
    satisfying
    \begin{align} 
	   {\bf x}_\text{eq} = {\bf x}_\text{eq} 
	   + \int^{\tau = t+\Delta t}_{\tau = t} {\bf f}[{\bf x}(\tau)] d\tau
    \end{align}
    for all $t \in [0, \infty)$, where ${\bf f}[{\bf x}(\tau)]$ is bounded by \eqref{eq:f_bound_L_and_M},
    is exponentially asymptotically stable
	if there exist positive constants, $r_0$, $k (\ge 1)$ and $\lambda$ such that
	\begin{align}
		\|{\boldsymbol \phi}(t, t_0, {\bf x}_0)\| \le
		k  \|{\bf x}_0\| e^{-\lambda(t-t_0)}
	\end{align}
	for all ${\bf x}_0 = {\bf x}(t_0) \in {\mathbb D}_0$ and $t \ge t_0 \ge 0$,
	where ${\mathbb D}_0 = \{ {\bf x} \in {\mathbb R}^n | \|{\bf x}\| \le r_0 \}$.
\end{definition}
\bigskip

Finding the maximum $r_0$ satisfying the exponential stability and the size of the domain of attraction is of high interest in system stability verification.

\bigskip

\begin{assumption}[Exponential Stable]
The nonlinear system given by 
\eqref{eq:x_t_x0_int_f_no_time_dep} with the discontinuous function bounded by \eqref{eq:f_bound_L_and_M} is assumed to be exponentially stable at the
equilibrium point, ${\bf x}_\text{eq}$, for all ${\bf x}_0 \in {\mathbb D}_0$.
\end{assumption}

\bigskip

\begin{theorem}[Exponential Bounds]
    With the exponentially stable assumption, the following bound is satisfied:
    \begin{align} \label{eq:exponential_bound}
    \| {\boldsymbol\phi}_N(N \Delta t, t, {\bf x}) - {\boldsymbol\phi}_N(N \Delta t, t, {\bf y}) \|
    \le 2 k r_0 e^{-\lambda T}
    \end{align}
\end{theorem}
\emph{Proof: } By the triangle inequality,
\begin{align}
    &\| {\boldsymbol\phi}_N(N \Delta t, t, {\bf x}) - {\boldsymbol\phi}_N(N \Delta t, t, {\bf y}) \| \notag\\
    &\le \| {\boldsymbol\phi}_N(N \Delta t, t, {\bf x})\| + \|{\boldsymbol\phi}_N(N \Delta t, t, {\bf y}) \|
\end{align}
Due to the exponential stable assumption and the definition of ${\mathbb D}_0$,
the following inequalities are satisfied:
\begin{align}
    &\| {\boldsymbol\phi}_N(N \Delta t, t, {\bf x})\| + \|{\boldsymbol\phi}_N(N \Delta t, t, {\bf y}) \|
    \notag\\
    &\le
    k  \|{\bf x}\| e^{-\lambda T} + k  \|{\bf y}\| e^{-\lambda T}
    \le
    2 k r_0 e^{-\lambda T}
\end{align}
Hence, the inequality, \eqref{eq:exponential_bound}, is satisfied.
\begin{flushright}
$\blacksquare$    
\end{flushright}
\bigskip

\begin{theorem}[Square-root Bound] \label{theorem:square_root_bound}
The solution of the exponential stable nonlinear systems satisfies the following inequality:
\begin{align} \label{eq:multiplied_bound}
 &\| {\boldsymbol\phi}_N(N \Delta t, t, {\bf x}) - {\boldsymbol\phi}_N(N \Delta t, t, {\bf y}) \|
 \notag\\
 &\le \sqrt{2 k r_0 e^{-\lambda T} a \| {\bf x} - {\bf y} \| + 2 k r_0 e^{-\lambda T} b}
\end{align} 
\end{theorem}
\emph{Proof:} Multiplying the bound in \eqref{eq:phi_bound_for_long_time_T} and the exponential bound in \eqref{eq:exponential_bound} and square-root both sides produces the inequality. $\blacksquare$      

\bigskip

\begin{figure}
    \centering
    \includegraphics[width=0.85\linewidth]{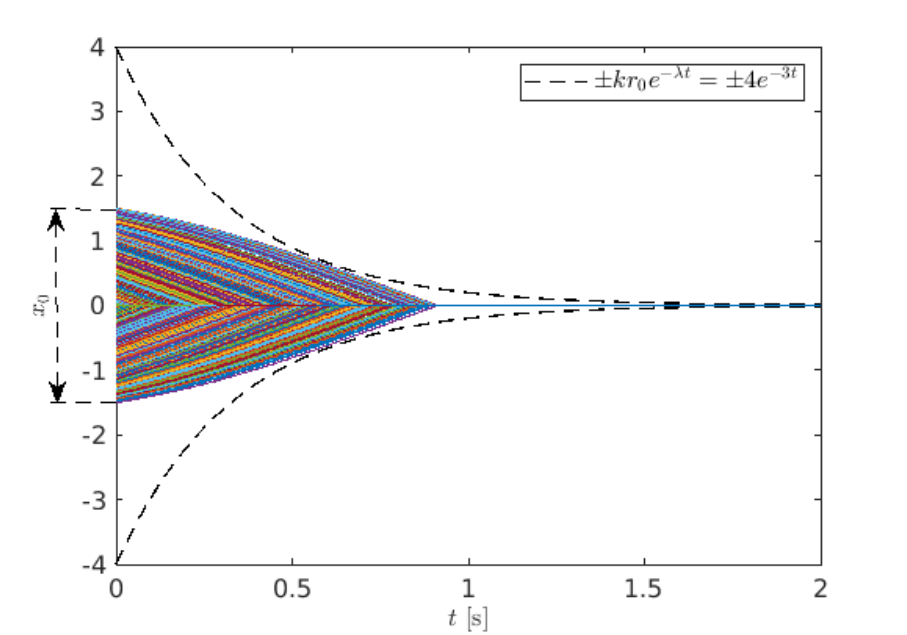}
    \caption{The exponential bounds and the 1000 Monte-Carlo simulations}
    \label{fig:example_01_x_bound}
\end{figure}

\begin{figure}
    \centering
    \includegraphics[width=0.95\linewidth]{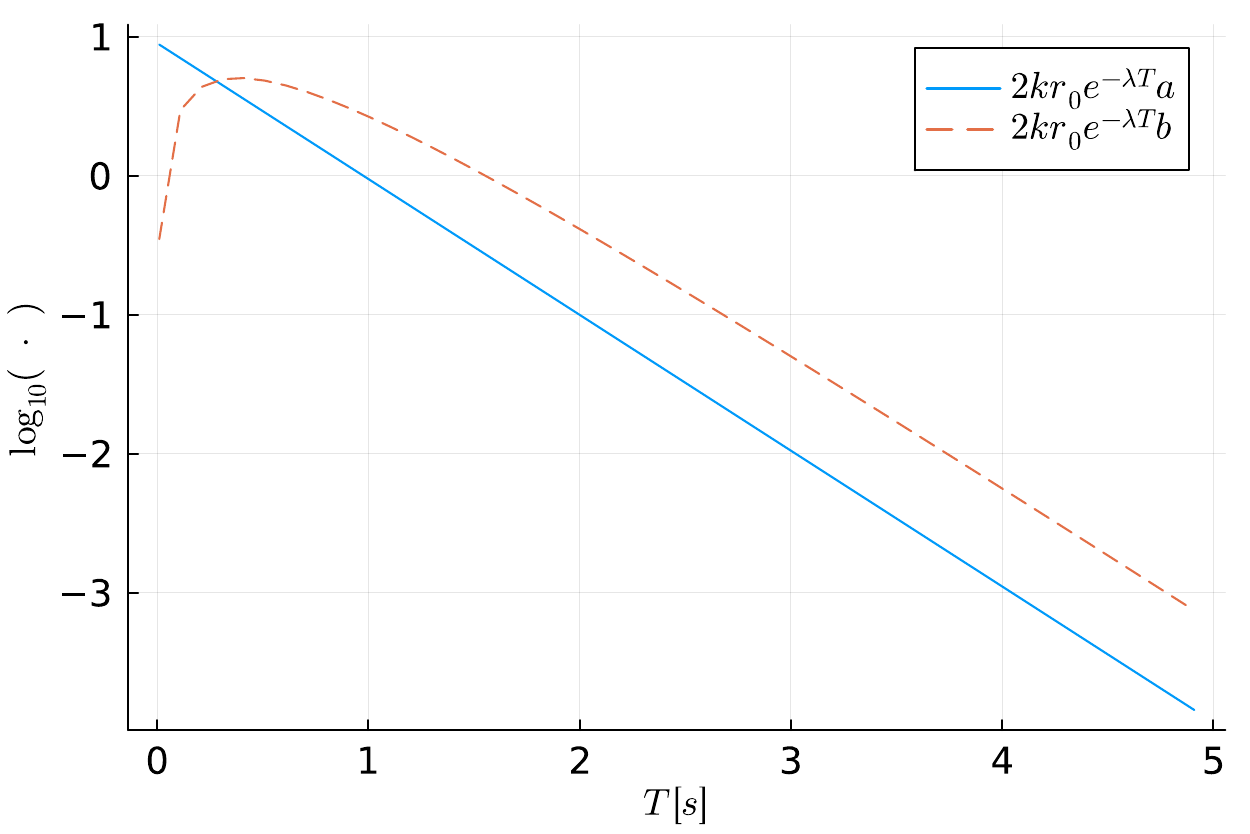}
    \caption{$\log_{10}(2 k r_0 e^{-\lambda T} a)$ and $\log_{10}(2 k r_0 e^{-\lambda T} b)$ with respect to $T$}
    \label{fig:square_a_b_example}
\end{figure}

\begin{example} \label{example:exponential_const}
For the nonlinear system given in Example \ref{example:sgn_cubic},
it is identified that $k = 8/3$, $r_0 = 3/2$ and $\lambda = 3$. 
The bounds for $|x_0|$ approaching $r_0$ with 1,000 simulation time histories are shown in Figure \ref{fig:example_01_x_bound}.
Using the bound obtained in Example \ref{example:sgn_cubic_a_b_bounds} with respect to $T$, the values of the square-root bound in \eqref{eq:multiplied_bound} are shown in
Figure \ref{fig:square_a_b_example}.
Both values become smaller than 0.001 around $T = 5$s, where $N=500$.
\end{example}
\bigskip


\begin{remark}[Choice of Simulation Time Interval $T$]
$T$ determines the time length of the numerical simulator. The longer $T$ requires a longer simulation time and the shorter $T$ results in larger values for the bound. The larger bound values require tighter samples, i.e., more samples, to check the stability conditions.
\end{remark}
\bigskip

\begin{definition}[Energy \& Energy Integral]
    The system energy, $E[{\boldsymbol{\phi}}_N(\tau, t, {\bf x}(t))]$, is defined by
    \begin{align}
        E[{\boldsymbol{\phi}}_N(\tau, t, {\bf x}(t))] = \dfrac{1}{2}
	    {\boldsymbol \phi}_N[\tau, t, {\bf x}(t)]^T P~ {\boldsymbol \phi}_N[\tau,t, {\bf x}(t)]
    \end{align}
    where $P$ is an $n\times n$ positive-definite matrix,
    and the energy integral function, $V[t,{\bf x}(t)]$, is defined by
    \begin{align}
        V[t,{\bf x}(t)] = \int^{t+T}_t E[{\boldsymbol{\phi}}_N(\tau, t, {\bf x}(t))]\, d\tau 
    \end{align}
\end{definition}

\bigskip

\begin{theorem}[Bounds of Energy \& Energy Integral]
    As the system is exponentially stable, the energy is bounded by
    \begin{align}
        E[{\boldsymbol{\phi}}_N(\tau, t, {\bf x}(t))] 
        \le \dfrac{k_E}{2} k^2  e^{-2\lambda(\tau - t)} \|{\bf x}(t)\|^2
    \end{align}
    where $k_E$ is the maximum eigenvalue of $P$,
    and the energy integral is bounded by
    \begin{align}
        V[t,{\bf x}(t)]
        \le  T \dfrac{k_E}{2} k^2  e^{-2\lambda(\tau - t)} \|{\bf x}(t)\|^2
    \end{align}
\end{theorem}
\emph{Proof:} The proof is trivial and omitted. $\blacksquare$

\bigskip

\begin{example} \label{example:energy_bound}
For the nonlinear system given in Example \ref{example:sgn_cubic} with the constants
identified in Example \ref{example:exponential_const}, let the energy
be given by
\begin{align*}
    E(\tau, 0, x_0) &= \dfrac{1}{2} \phi^2_{150}(\tau, 0, x_0) \notag\\
    &\le \dfrac{1}{2} \left(\dfrac{8}{3}\right)^2 e^{-6 \tau} x^2_0
    =  \dfrac{32}{9} e^{-6 \tau} x^2_0
\end{align*}
where $k_E = 1$ and $x_0$ is the initial state in ${\mathbb D}_0$,
and $V(t,x_0)$ is bounded by $T E(\tau, 0, x_0)$.
\end{example}

\bigskip

\begin{theorem}[Energy slope bound]
The energy function difference is bounded by
\begin{align} \label{eq:energy_slope_bound}
	&\left| E[{\boldsymbol{\phi}}_N(\tau, t, {\bf x})] 
    -  E[{\boldsymbol{\phi}}_N(\tau, t, {\bf y})]\right| 
    \notag\\
    &\le k_E \sqrt{2 k r_0 e^{-\lambda T} a \| {\bf x} - {\bf y}
    \| + 2 k r_0 e^{-\lambda T} b}
\end{align} 
\end{theorem}
\emph{Proof:} 
By the definition of the energy function, its slope, i.e., $\partial E/\partial {\boldsymbol\phi}$
is bounded by $k_E$. Hence, the difference is also bounded by
\begin{align}
	&\left| E[{\boldsymbol{\phi}}_N(\tau, t, {\bf x})] 
    -  E[{\boldsymbol{\phi}}_N(\tau, t, {\bf y})]\right| 
    \notag\\
    &\le k_E \left|{\boldsymbol{\phi}}_N(\tau, t, {\bf x}) 
    -  {\boldsymbol{\phi}}_N(\tau, t, {\bf y})\right|  
\end{align}
and due to the bound given by Theorem \ref{theorem:square_root_bound},
\begin{align}
	&k_E \| {\boldsymbol{\phi}}_N(\tau, t, {\bf x}) 
    - {\boldsymbol{\phi}}_N(\tau, t, {\bf y}) \|
	\notag\\
    &\le k_E \sqrt{2 k r_0 e^{-\lambda T} a \| {\bf x} - {\bf y} \| + 2 k r_0 e^{-\lambda T} b}
\end{align}
\begin{flushright}
$\blacksquare$    
\end{flushright}

\bigskip

\begin{example} \label{example:energy_diff_bound}
For the nonlinear system given in Example \ref{example:sgn_cubic} with the energy defined
in Example \ref{example:energy_bound}, the energy slope bound is obtained as
\begin{align*}
    &\left| E[{\boldsymbol{\phi}}_{300}(3, 0, x)] 
    -  E[{\boldsymbol{\phi}}_{300}(3, 0, y)]\right| \notag\\
    &\le \sqrt{ 8 e^{-9}\times 9.49 | x - y | + 8 e^{-9}\times 45.27} \notag\\
    &\approx \sqrt{ 0.0094 | x - y | + 0.0447}
\end{align*}
where $N=300$ and $\Delta t = 0.01$s.
\end{example}

\bigskip

\begin{theorem}[$\delta$-sampling] \label{theorem:delta_sampling}
Choose ${\bf x}_\delta$ in ${\mathbb S}_\delta$ such that
\begin{align}
    \|{\bf x} - {\bf x}_\delta \| \le \delta
\end{align}
where ${\mathbb S}_\delta$ is a finite subset of ${\mathbb S}$,
${\mathbb S}$ is a subset of ${\mathbb R}^n$. Then,
\begin{align}
	&\left| E[{\boldsymbol{\phi}}_N(\tau, t, {\bf x})] 
    -  E[{\boldsymbol{\phi}}_N(\tau, t, {\bf x}_\delta)]\right| 
    \notag\\
    &\le k_E \sqrt{2 k r_0 e^{-\lambda T} a \| {\bf x} - {\bf x}_\delta
    \| + 2 k r_0 e^{-\lambda T} b}
    \notag\\
    &\le 
    k_E \sqrt{2 k r_0 e^{-\lambda T} a \delta
    + 2 k r_0 e^{-\lambda T} b}
\end{align}
\end{theorem}

\bigskip

\begin{definition}[Forward Invariant Set]
    Let ${\mathbb S}$ be the set of ${\bf x}$, whose corresponding energy is less
    than $\ell$, i.e.,
    \begin{align}
        {\mathbb S} = \left\{ {\bf x} | E({\bf x}) \le \ell \right\}
    \end{align}
    and the following inequality is satisfied for all ${\bf x}$ in ${\mathbb S}$
    \begin{align}
        E[{\boldsymbol\phi}_N(T, 0, {\bf x})] \le \ell, 
    \end{align}
    then the set ${\mathbb S}$ is \emph{forward invariant}.
\end{definition}
\bigskip

\begin{figure}
\centering
\includegraphics[width=0.95\linewidth,trim={3cm 9.6cm 4cm 9.6cm},clip]{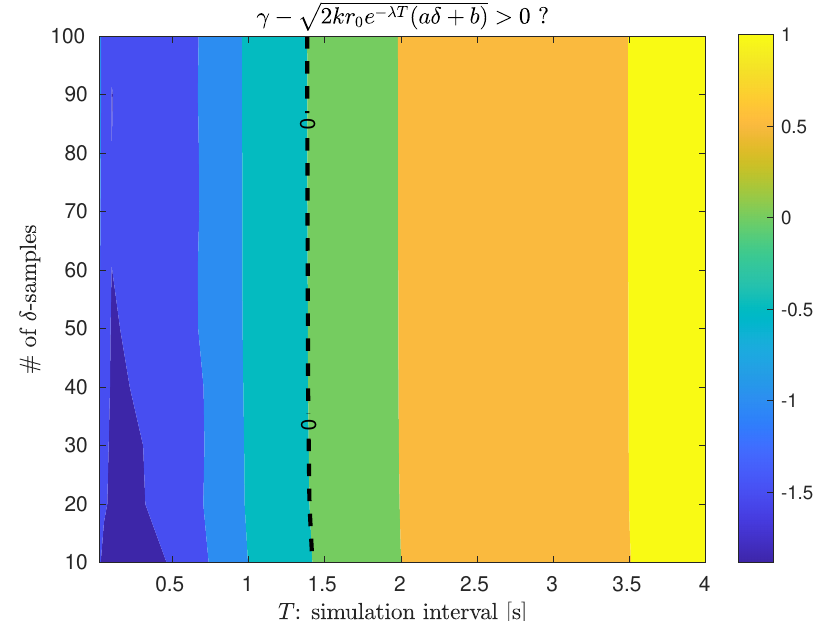}
\caption{The inequality condition values for the ranges of the
number of $x_\delta$ samplings and the simulation time length, $T$}
\label{fig:T_N_gama_ineq}
\end{figure}
\begin{theorem}[Verification of Forward Invariant] 
    For all ${\bf x}_\delta$ in ${\mathbb S}_\delta$, if there exists a positive
    real $\gamma$ such that
    \begin{align} \label{eq:gamma_cond_01}
        E[{\boldsymbol\phi}_N(T, 0, {\bf x}_\delta )] \le \ell - \gamma 
    \end{align}
    and
    \begin{align} \label{eq:gamma_cond_02}
        \sqrt{2 k r_0 e^{-\lambda T} a \delta
    + 2 k r_0 e^{-\lambda T} b} \le \gamma, 
    \end{align}
    then ${\mathbb S}$ is forward invariant \cite{10.1007/978-3-319-12307-3_37}.
\end{theorem}
\emph{Proof:} Prove it by contradiction as in \cite{10.1007/978-3-319-12307-3_37}.
Assume \eqref{eq:gamma_cond_01} and \eqref{eq:gamma_cond_02} are satisfied
but there exits ${\bf x}^*$ in ${\mathbb S}$ such that
\begin{align} \label{eq:bad_x_star_in_S}
    E[{\boldsymbol\phi}_N(T, 0, {\bf x}^*)] > \ell \rightarrow -E[{\boldsymbol\phi}_N(T, 0, {\bf x}^*)] < -\ell
\end{align}
Add \eqref{eq:gamma_cond_01} and \eqref{eq:bad_x_star_in_S}
\begin{align}
    E[{\boldsymbol\phi}_N(T, 0, {\bf x}_\delta )] - E[{\boldsymbol\phi}_N(T, 0, {\bf x}^*)] < -\gamma < 0
\end{align}
Hence,
\begin{align}
    \| E[{\boldsymbol\phi}_N(T, 0, {\bf x}_\delta )] - E[{\boldsymbol\phi}_N(T, 0, {\bf x}^*)] \| > \gamma
\end{align}
Because of the energy slope bound in \eqref{eq:energy_slope_bound}
\begin{align}
    \gamma 
    &<
    \| E[{\boldsymbol\phi}_N(T, 0, {\bf x}_\delta )] - E[{\boldsymbol\phi}_N(T, 0, {\bf x}^*)] \|
    \notag \\
    &\le
    k_E \sqrt{2 k r_0 e^{-\lambda T} a \| {\bf x}_\delta - {\bf x}^*
    \| + 2 k r_0 e^{-\lambda T} b}
\end{align}
As ${\bf x}_\delta$ belongs to the $\delta$-sampling set, ${\mathbb S}_\delta$ in Theorem \ref{theorem:delta_sampling},
\begin{align}
    \gamma 
    &< 
    \| E[{\boldsymbol\phi}_N(T, 0, {\bf x}_\delta )] - E[{\boldsymbol\phi}_N(T, 0, {\bf x}^*)] \|
    \notag \\
    &\le
    k_E \sqrt{2 k r_0 e^{-\lambda T} a \| {\bf x}_\delta - {\bf x}^*
    \| + 2 k r_0 e^{-\lambda T} b} \notag \\
    &\le
    k_E \sqrt{2 k r_0 e^{-\lambda T} a \delta + 2 k r_0 e^{-\lambda T} b}
\end{align}
The right-most term is bounded by \eqref{eq:gamma_cond_02} and finally,
\begin{align}
    \gamma 
    < 
    \| E[{\boldsymbol\phi}_N(T, 0, {\bf x}_\delta )] - E[{\boldsymbol\phi}_N(T, 0, {\bf x}^*)] \|
    \le \gamma
\end{align}
The inequality contradicts. Hence, the assumption of the existence of ${\bf x}^*$ is incorrect.
\begin{flushright}
$\blacksquare$    
\end{flushright}
\bigskip

\begin{algorithm}
	\caption{Stability Verification with $\delta$-Samples}
\label{ALG:stability_check}
\begin{algorithmic}[1]
    \State Set $\Delta t$, $N$, $\delta$, $k$, $\lambda$ and $\ell$ 
    \State Generate ${\bf x}_\delta \in {\mathbb S}_\delta \subset {\mathbb S}$
    \While {True} 
    	\State Run Simulator for each ${\bf x}_\delta$
        \State Calculate 
        $\gamma = \ell - \max E[{\boldsymbol\phi}_N(T,0,{\bf x}_\delta)]$ 
        \If {$\gamma > 0$} 
            \If {\eqref{eq:gamma_cond_02} satisfies} 
                \If {all ${\bf x}_\delta$ checked}
                    ${\mathbb S}$ is forward-invariant.
                \Else { go to the next sample}
                \EndIf
            \Else {} reduce $\delta$ and go to ${\bf x}_\delta$ generation
            \EndIf
            \Else {} adjust $N$, $\delta$, $k$, $\lambda$ and $\ell$ and start over 
        \EndIf
    \EndWhile
\end{algorithmic} 
\end{algorithm}
\begin{example} \label{example:gamma_bound}
For the nonlinear system given in Example \ref{example:sgn_cubic} with the energy difference bound
in Example \ref{example:energy_diff_bound}, the inequality for $\delta$ and $\gamma$ must satisfy as follows:
\begin{align*}
  \sqrt{0.0094 | x - y | + 0.0447} \le \gamma
\end{align*}
The inequality provides the minimum $\gamma$ bound equal to 0.0447, where
$\delta$ is equal to zero corresponding to the infinitely many samples.
As $(\ell-\gamma)$ in \eqref{eq:gamma_cond_01} must be positive,
the smaller minimum $\gamma$ increases the chance that the inequality
in \eqref{eq:gamma_cond_01} satisfies with a positive value of $\delta$, 
i.e., a finite number of samples. Change $N=400$, i.e., $T=4$s, then
the following inequality is calculated
\begin{align*}
  \sqrt{ 0.0009 \delta + 0.005} \le \gamma
\end{align*}
\end{example}
and the lower bound of the minimum $\gamma$ is reduced to 0.005.
\bigskip

Algorithm \ref{ALG:stability_check} summarizes the stability
check using a finite number of samples.
\bigskip

\begin{remark}
Overestimating $k$, which is related to overshoots of the response,  
in the Algorithm is allowed 
with the price that longer simulation time interval, i.e.,
larger, $N$, would need to satisfy the inequalities.
On the other hand, $\lambda$ must be underestimated.
\end{remark}
\bigskip

\begin{example}
    For Example \ref{example:sgn_cubic}, each variable in 
    Algorithm \ref{ALG:stability_check} is given by
    $\Delta t = 0.01$, $k = 8/3$, $\lambda = 3$ and 
    $\ell = (k_E r_0^2)/2$, where $k_E = 1.0$ and $r_0 = 3/2$.
    $N$ varies from 10 to 100.
    $T$ is determined by $N\Delta t$.
    For each $N$, $N_\text{samp}$ number of $x_\delta$ samples
    are obtained in $-r_0 < x_\delta < r_0$, while the maximum
    distance between the samples is kept less than $\delta/2$.
    The $\gamma$ inequality condition for each combination of 
    the number of $x_\delta$ samples and $T$ is shown
    in Figure \ref{fig:T_N_gama_ineq}.
    For this example, when the stability inequality condition
    is violated, it is better to increase $T$ instead of decreasing
    $\delta$.
\end{example}

\section{Conclusions \& Future Works}
We present a stability verification method providing the deterministic
stability assurance of dynamical systems implemented as
high-fidelity numerical simulators, which may include hard nonlinear
components such as discontinuous jumps in the states and magnitude/speed
constraints, and simulation-based control design algorithms 
such as reinforcement learning,
which does not provide a stability guarantee by the design procedures.
For each specific real-world application, there would be abundant room
to improve the proposed algorithm in terms of a parallelization of the 
algorithm, an efficient sampling and a better estimation of
$k$ and $\lambda$ leading to allow the larger $\delta$ and/or
the smaller $\gamma$. For example,
\cite{8267231} provides the way of efficient sampling.

\section*{Acknowledgement}
This research was supported by Unmanned Vehicles Core Technology Research and Development Program (No.\,2020M3C1C1A0108316111) 
through the National Research Foundation of Korea (NRF) and 
Unmanned Vehicle Advanced Research Center(UVARC) funded by the Ministry of Science and ICT, Republic of Korea.

\bibliographystyle{IEEEtran}
\bibliography{IEEEabrv,conv_Lyp_stab_verify}

\begin{thebibliography}{10}
\providecommand{\url}[1]{#1}
\csname url@samestyle\endcsname
\providecommand{\newblock}{\relax}
\providecommand{\bibinfo}[2]{#2}
\providecommand{\BIBentrySTDinterwordspacing}{\spaceskip=0pt\relax}
\providecommand{\BIBentryALTinterwordstretchfactor}{4}
\providecommand{\BIBentryALTinterwordspacing}{\spaceskip=\fontdimen2\font plus
\BIBentryALTinterwordstretchfactor\fontdimen3\font minus
  \fontdimen4\font\relax}
\providecommand{\BIBforeignlanguage}[2]{{%
\expandafter\ifx\csname l@#1\endcsname\relax
\typeout{** WARNING: IEEEtran.bst: No hyphenation pattern has been}%
\typeout{** loaded for the language `#1'. Using the pattern for}%
\typeout{** the default language instead.}%
\else
\language=\csname l@#1\endcsname
\fi
#2}}
\providecommand{\BIBdecl}{\relax}
\BIBdecl

\bibitem{1102565}
N.~Lehtomaki, N.~Sandell, and M.~Athans, ``Robustness results in
  linear-quadratic gaussian based multivariable control designs,'' \emph{IEEE
  Transactions on Automatic Control}, vol.~26, no.~1, pp. 75--93, 1981.

\bibitem{29425}
J.~Doyle, K.~Glover, P.~Khargonekar, and B.~Francis, ``State-space solutions to
  standard {$H_2$} and {$H_\infty$} control problems,'' \emph{IEEE Transactions
  on Automatic Control}, vol.~34, no.~8, pp. 831--847, 1989.

\bibitem{1101446}
V.~Utkin, ``Variable structure systems with sliding modes,'' \emph{IEEE
  Transactions on Automatic Control}, vol.~22, no.~2, pp. 212--222, 1977.

\bibitem{morari1999model}
M.~Morari and J.~H. Lee, ``Model predictive control: past, present and
  future,'' \emph{Computers \& chemical engineering}, vol.~23, no. 4-5, pp.
  667--682, 1999.

\bibitem{7741019}
J.~Kapinski, J.~V. Deshmukh, X.~Jin, H.~Ito, and K.~Butts, ``Simulation-based
  approaches for verification of embedded control systems: An overview of
  traditional and advanced modeling, testing, and verification techniques,''
  \emph{IEEE Control Systems Magazine}, vol.~36, no.~6, pp. 45--64, 2016.

\bibitem{5340639}
S.~Karimi, P.~Poure, and S.~Saadate, ``A {HIL}-based reconfigurable platform
  for design, implementation, and verification of electrical system digital
  controllers,'' \emph{IEEE Transactions on Industrial Electronics}, vol.~57,
  no.~4, pp. 1226--1236, 2010.

\bibitem{8723561}
S.~Chen, Y.~Chen, S.~Zhang, and N.~Zheng, ``A novel integrated simulation and
  testing platform for self-driving cars with hardware in the loop,''
  \emph{IEEE Transactions on Intelligent Vehicles}, vol.~4, no.~3, pp.
  425--436, 2019.

\bibitem{lee2021development}
D.~H. Lee, C.-J. Kim, and S.~H. Lee, ``Development of unified high-fidelity
  flight dynamic modeling technique for unmanned compound aircraft,''
  \emph{International Journal of Aerospace Engineering}, vol. 2021, pp. 1--23,
  2021.

\bibitem{foster2017high}
J.~V. Foster and D.~Hartman, ``High-fidelity multi-rotor unmanned aircraft
  system ({UAS}) simulation development for trajectory prediction under
  off-nominal flight dynamics,'' in \emph{17th AIAA Aviation Technology,
  Integration, and Operations Conference}, 2017, p. 3271.

\bibitem{hieb2013creating}
B.~Hieb, ``Creating a high-fidelity model of an electric motor for control
  system design and verification,'' \emph{Technical Article Published by The
  MathWorks}, 2013.

\bibitem{sutton1999reinforcement}
R.~S. Sutton, A.~G. Barto \emph{et~al.}, ``Reinforcement learning,''
  \emph{Journal of Cognitive Neuroscience}, vol.~11, no.~1, pp. 126--134, 1999.

\bibitem{lillicrap2015continuous}
T.~P. Lillicrap, J.~J. Hunt, A.~Pritzel, N.~Heess, T.~Erez, Y.~Tassa,
  D.~Silver, and D.~Wierstra, ``Continuous control with deep reinforcement
  learning,'' \emph{arXiv preprint arXiv:1509.02971}, 2015.

\bibitem{vukobratovic2003dynamics}
M.~Vukobratovic, V.~Potkonjak, and V.~Matijevic, \emph{Dynamics of robots with
  contact tasks}.\hskip 1em plus 0.5em minus 0.4em\relax Springer Science \&
  Business Media, 2003, vol.~26.

\bibitem{khalil2015nonlinear}
H.~K. Khalil, \emph{Nonlinear control}.\hskip 1em plus 0.5em minus 0.4em\relax
  Pearson New York, 2015, vol. 406.

\bibitem{trumpf2010converse}
J.~Trumpf and R.~Mahony, ``A converse liapunov theorem for uniformly locally
  exponentially stable systems admitting {Carath{\`e}odory} solutions,''
  \emph{IFAC Proceedings Volumes}, vol.~43, no.~14, pp. 1374--1378, 2010.

\bibitem{4518905}
J.~Cortes, ``Discontinuous dynamical systems,'' \emph{IEEE Control Systems
  Magazine}, vol.~28, no.~3, pp. 36--73, 2008.

\bibitem{10.1007/978-3-319-12307-3_37}
J.~Kapinski and J.~Deshmukh, ``Discovering forward invariant sets for nonlinear
  dynamical systems,'' in \emph{Interdisciplinary Topics in Applied
  Mathematics, Modeling and Computational Science}, M.~G. Cojocaru, I.~S.
  Kotsireas, R.~N. Makarov, R.~V.~N. Melnik, and H.~Shodiev, Eds.\hskip 1em
  plus 0.5em minus 0.4em\relax Cham: Springer International Publishing, 2015,
  pp. 259--264.

\bibitem{8267231}
R.~Bobiti and M.~Lazar, ``Automated-sampling-based stability verification and
  doa estimation for nonlinear systems,'' \emph{IEEE Transactions on Automatic
  Control}, vol.~63, no.~11, pp. 3659--3674, 2018.

\end{thebibliography}

\end{document}